\documentclass[12pt,pra,aps,amssymb,amsmath,showpacs,tightenlines]{revtex4}
\usepackage{dsfont}

\newcommand{\half}{\mbox{$\textstyle \frac{1}{2}$}}
\newcommand{\quat}{\mbox{$\textstyle \frac{1}{4}$}}

\newcommand{\re}{\mbox{$\rm e$}}
\newcommand{\ri}{\mbox{$\rm i$}}

\begin{document}
\rightline{preprint LA-UR-07-0447}

\newtheorem{guess}{Proposition}

\title{Geometry of PT-symmetric quantum mechanics}

\author{Carl~M.~Bender$^{1}$\footnote{Permanent
address: Department of Physics, Washington University, St. Louis MO
63130, USA}, Dorje~C.~Brody$^{2}$, \\ Lane~P.~Hughston,$^{3}$ and
Bernhard~K.~Meister$^{4}$}

\affiliation{${}^{1}$Center for Nonlinear Studies, Los Alamos
National Laboratory, Los Alamos, NM 87545, USA \\ ${}^{2}$Department
of Mathematics, Imperial College, London SW7 2BZ, UK \\
${}^{3}$Department of Mathematics, King's College London, London
WC2R 2LS, UK \\ ${}^{4}$Department of Physics, Renmin University of
China, Beijing 100872, China}

\date{\today}

\begin{abstract}
Recently, much research has been carried out on Hamiltonians that
are not Hermitian but are symmetric under space-time reflection,
that is, Hamiltonians that exhibit $\mathcal{PT}$ symmetry.
Investigations of the Sturm-Liouville eigenvalue problem associated
with such Hamiltonians have shown that in many cases the entire
energy spectrum is real and positive and that the eigenfunctions
form an orthogonal and complete basis. Furthermore, the quantum
theories determined by such Hamiltonians have been shown to be
consistent in the sense that the probabilities are positive and the
dynamical trajectories are unitary. However, the geometrical
structures that underlie quantum theories formulated in terms of
such Hamiltonians have hitherto not been fully understood. This
paper studies in detail the geometric properties of a Hilbert space
endowed with a parity structure and analyses the characteristics of
a ${\cal PT}$-symmetric Hamiltonian and its eigenstates. A canonical
relationship between a ${\cal PT}$-symmetric operator and a
Hermitian operator is established. It is shown that the quadratic
form corresponding to the parity operator, in particular, gives rise
to a natural partition of the Hilbert space into two halves
corresponding to states having positive and negative ${\cal PT}$
norm. The indefiniteness of the norm can be circumvented by
introducing a symmetry operator ${\mathcal C}$ that defines a
positive definite inner product by means of a ${\cal CPT}$
conjugation operation.
\end{abstract}


\maketitle

\section{Introduction}
\label{s1}

In standard quantum mechanics it is assumed that the Hamiltonian $H$
is Hermitian. This requirement ensures that the spectrum of $H$ is
real. However, in the past decade many researchers have investigated
the consequences of replacing the mathematical requirement of
Hermiticity by a more directly physical discrete space-time
reflection symmetry known as $\mathcal{PT}$ invariance, where
$\mathcal P$ is the parity reflection operator and $\mathcal T$ is
the time reversal operator (Znojil 2004, 2005, 2006, Bender 2005,
Geyer {\it et al}. 2006, Bender 2007). In particular, if
$\mathcal{PT}$ symmetry is not broken, that is, if the
eigenfunctions of the Hamiltonian $H$ are simultaneously
eigenfunctions of the $\mathcal{PT}$ operator, then the spectrum of
the Hamiltonian is entirely real (Bender \& Boettcher 1998, Bender
\textit{et al}. 1999, Dorey \textit{et al}. 2001a, 2001b, 2007).
Furthermore, if a Hilbert space is constructed in terms of an
appropriate inner product, then a quantum theory described by a
$\mathcal{PT}$-symmetric Hamiltonian exhibits all the desired
physical features (Bender \textit{et al}. 2002b, Mostafazadeh 2002).

Hermiticity is a strong condition. Not only does it guarantee the
reality of the spectrum, it also generates unitary time evolution.
In addition, Hermiticity ties in with a positive definite inner
product, which leads to the usual probabilistic interpretation of
quantum mechanics. These three results follow naturally from the
assumption of Hermiticity. The condition of $\mathcal{PT}$ symmetry
is a distinct requirement from Hermiticity. Nevertheless, given the
observation that $\mathcal{PT}$-symmetric operators may possess real
eigenvalues, it is legitimate to ask (a) whether a physically viable
quantum theory can be formulated when we replace the Hermiticity
condition with the requirement of space-time reflection symmetry,
and (b) whether this new formulation may lead to new physical
predictions. Indeed, investigations over the past nine years have
shown that by introducing a new symmetry operator denoted as ${\cal
C}$, a Hilbert space with a positive-definite inner product can be
constructed upon which $\mathcal{PT}$-symmetric Hamiltonians act as
self-adjoint operators. As a consequence, consistent quantum
theories can be formulated via Hamiltonians that possess space-time
reflection symmetry but are not Hermitian in the conventional sense.

While many examples of $\mathcal{PT}$-symmetric quantum theories
have been analysed in the literature, some of the basic mathematical
structures of the theory, such as the geometry of the underlying
real Hilbert space in which $\mathcal{PT}$-symmetric quantum
theories are defined, have not been fully characterised. The present
paper addresses this question by clarifying various mathematical
structures of the underlying Hilbert space. For the purpose of
constructing a viable quantum theory, we need to consider a
framework sufficiently general to admit both the standard theory
with a Hermitian Hamiltonian as well as extensions of the standard
theory. Thus, we discuss in Section \ref{s2} and Section \ref{s3}
the geometrical structures of the underlying real Hilbert space and
the role of the observables in conventional quantum mechanics.

In Section \ref{s4} and Section \ref{s5} we compare the structures
described in Sections \ref{s2} and \ref{s3} with the corresponding
structures in the quantum theory symmetric under space-time
reflection. It is known that the requirement of $\mathcal{PT}$
symmetry alone on the Hamiltonian leads to a state space with an
indefinite metric. The important observation we make is that the
parity operator associated with space reflection plays the role of
an indefinite metric, while the complex structure $J$ of standard
quantum mechanics is unaltered in the ${\cal PT}$-symmetric theory.
This is an attractive feature of ${\cal PT}$-symmetric quantum
theory from the point of view of complex analysis. We show in
Proposition~\ref{prop1} that the squared ${\cal PT}$ norm of a state
is expressible as a difference of the squared standard Dirac norms
of the positive and negative parity parts of the state. Section
\ref{s5} also discusses observables. See Mostafazadeh \& Batal
(2004), Mostafazadeh (2005), and Jones (2005) for previous work on
observables in ${\cal PT}$-symmetric quantum theories.

In Section \ref{s6} we analyse the properties of Hamiltonian
operators that are symmetric under space-time reflection. It is
shown in Proposition~\ref{prop2} that any such Hamiltonian is
necessarily expressed as a product of the parity structure and a
Hermitian quadratic form. This leads to an alternative way of
understanding the reality of the spectrum of such Hamiltonians, as
established in Proposition~\ref{prop3}, showing that the energy
eigenvalues are necessarily real if the corresponding eigenvectors
have nonvanishing ${\cal PT}$ norms. It is known in the literature
that the eigenvalues of ${\cal PT}$-symmetric Hamiltonians occur
either as real numbers or as complex conjugate pairs. This is shown
in Proposition~\ref{prop4}. A sufficient condition for the
orthogonality of the eigenstates is then established in
Proposition~\ref{prop5}. In Section \ref{s7} we define in
geometrical terms a reflection operator ${\cal C}$ whose
mathematical structure resembles that of a charge operator. This
symmetry operator allows us to construct an alternative inner
product on the vector space spanned by the eigenfunctions of the
${\cal PT}$-symmetric Hamiltonian in terms of ${\cal
CPT}$-conjugation, thus eliminating states having negative norms. As
a consequence, a consistent probabilistic interpretation can be
assigned to quantum theories described by ${\cal PT}$-symmetric
Hamiltonians. To construct the operator ${\cal C}$ we establish in
Proposition~\ref{prop6} that the eigenfunction associated with a
real eigenvalue of a ${\cal PT}$-symmetric Hamiltonian is either
real or purely imaginary, depending on its parity type. To
illustrate these ideas a system of ${\cal PT}$-symmetric
spin-$\frac{1}{2}$ particles is presented in Section \ref{s8}.

\section{Hermitian quantum mechanics}
\label{s2}

Our ultimate objective is to determine the geometric structure of
$\mathcal{PT}$-symmetric quantum theory. With this in mind we show
in this section how to formulate the geometric structure of standard
quantum mechanics. In Sections \ref{s4} and \ref{s5} we clarify the
similarities and differences between the two formalisms. In standard
quantum theory Hermitian operators have a dual role as physical
observables and as the generators of the dynamics. To understand the
relation between these roles it is useful to present quantum
mechanics in terms of a primitive underlying even-dimensional {\sl
real} Hilbert space ${\mathcal H}$ rather than the complex Hilbert
space with respect to which it is usually formulated. We will see
that by introducing certain structures on ${\mathcal H}$ we arrive
at standard quantum theory. Then by considering an alternative set
of structures on ${\mathcal H}$ we arrive at
$\mathcal{PT}$-symmetric quantum theory, and the relationship
between the two theories becomes clear from a geometric perspective.

Using a standard index notation (see, for example, Geroch 1971,
Gibbons $\&$ Pohle 1993, Brody $\&$ Hughston 1998, 1999 and
references cited therein) we let the real vector $\xi^a$ denote a
typical element of ${\cal H}$. The real Hilbert space ${\cal H}$ is
to be regarded as coming equipped with a positive definite quadratic
form $g_{ab}$ satisfying $g_{ab}=g_{ba}$, with respect to which the
squared norm of the vector $\xi^a$ is given by $g_{ab}
\xi^{a}\xi^{b}$. Then if $\xi^a$ and $\eta^a$ are a pair of elements
of ${\cal H}$, we define their inner product by $g_{ab}\xi^{a}\eta^{b}$.

One can only recover the familiar apparatus of standard quantum
mechanics if we further require that ${\cal H}$ also be endowed with
a compatible complex structure. By a \textit{complex structure} we
mean a real tensor $J^a_{\ b}$ satisfying the following condition:
\begin{eqnarray}
J^a_{\ c}J^c_{\ b}=-\delta^a_{\ b}. \label{eq:3.1}
\end{eqnarray}
The complex structure is then said to be \textit{compatible} with
the symmetric quadratic form if $g_{ab}$ and $J^a_{\ b}$ commute;
that is,
\begin{eqnarray}
g_{ab}J^a_{\ c}J^b_{\ d}=g_{cd}. \label{eq:3.2}
\end{eqnarray}
If this condition holds, then $g_{ab}$ is said to be
\textit{$J$-invariant}. The compatibility condition is crucial in
the case of relativistic fields, where we insist that the creation
and annihilation operators satisfy canonical commutation relations
(Ashtekar $\&$ Magnon 1975).

A straightforward calculation shows that the $J$-invariance of
$g_{ab}$ implies that the tensor $\Omega_{ab}$ defined by
\begin{eqnarray}
\Omega_{ab}=g_{ac}J^c_{\ b} \label{eq:x.3}
\end{eqnarray}
is antisymmetric and nondegenerate, and thus defines a {\sl
symplectic structure} on ${\mathcal H}$. To see the antisymmetry of
$\Omega_{ab}$, we insert (\ref{eq:3.2}) into (\ref{eq:x.3}) to obtain
$\Omega_{ba}=g_{bc}J^c_{\ a}=g_{de}J^d_{\ b}J^e_{\ c}J^c_{\
a}=-g_{de}J^d_{\ b}\delta^e_{\ a}=-\Omega_{ab}$. The nondegeneracy
of $\Omega_{ab}$ becomes clear if we observe that the tensor
\begin{eqnarray}
\Omega^{ab} = g^{ac} g^{cd} \Omega_{cd}
\end{eqnarray}
acts as the required inverse. Indeed, we have
\begin{eqnarray}
\Omega^{ac} \Omega_{bc} &=& g^{ae} g^{cf} g_{eh}J^h_{\ f}
g_{bd}J^d_{\ c} \nonumber \\ &=& g_{bd}J^d_{\ c}J^a_{\ f} g^{cf}
\nonumber \\ &=& \delta^a_{\ b},
\end{eqnarray}
where in the last step we have used $J$-invariance $J^a_{\ c}J^b_{\
d}g^{cd}=g^{ab}$ of the tensor $g^{ab}$. The symplectic structure is
also compatible with $J^a_{\ b}$ in the sense that
\begin{eqnarray}
\Omega_{ab}J^a_{\ c}J^b_{\ d}=\Omega_{cd}. \label{eq:3.3}
\end{eqnarray}
This follows because $\Omega_{ab}J^a_{\ c}J^b_{\ d}=g_{ae}J^e_{\
b}J^a_{\ c}J^b_{\ d} = -g_{ae}\delta^e_{\ d}J^a_{\ c} =-
\Omega_{dc}=\Omega_{cd}$. We refer to relation ({\ref{eq:3.3}) by
saying that $\Omega_{ab}$ is $J$-invariant.

With this material at hand we can now elucidate the structure of
standard quantum mechanics in geometrical terms. The idea is to
endow the real Hilbert space ${\mathcal H}$ with a {\sl Hermitian
inner product}. If $\xi^{a}$ and $\eta^{a}$ are real Hilbert space
vectors, then their Hermitian inner product, which we write as
$\langle\eta| \xi\rangle$ using the Dirac notation, is given by the
complex expression
\begin{eqnarray}
\langle\eta|\xi\rangle = \half \eta^{a}(g_{ab} - \ri \Omega_{ab})
\xi^{b} . \label{eq:3.4}
\end{eqnarray}
Because the symplectic form $\Omega_{ab}$ is antisymmetric, it
follows that, apart from a factor of two, the Hermitian norm
agrees with the real Hilbertian norm:
\begin{eqnarray}
\langle\xi|\xi\rangle = \half g_{ab}\xi^{a}\xi^{b}.
\end{eqnarray}

To develop the theory further, we need to complexify the Hilbert
space ${\cal H}$, and we denote this complexified space by
${\mathcal H}_{\mathbb C}$. The elements of ${\mathcal H}_{\mathbb
C}$ are complex vectors of the form $\xi^{a}+ \ri \eta^{b}$, where
$\xi^{a}$ and $\eta^{b}$ are elements of the underlying real Hilbert
space ${\cal H}$.

With the aid of the complex structure, a real Hilbert space vector
$\xi^{a}$ can be decomposed into complex $J$-positive and
$J$-negative parts as follows:
\begin{eqnarray}
\xi^{a} = \xi^{a}_{+} + \xi^{a}_{-}, \label{eq:x.6}
\end{eqnarray}
where
\begin{eqnarray}
\xi^{a}_{+} = \half (\xi^{a}- \ri J^{a}_{\ b}\xi^{b}) \quad {\rm
and} \quad \xi^{a}_{-} = \half (\xi^{a}+ \ri J^{a}_{\ b}\xi^{b}) .
\label{eq:3.5}
\end{eqnarray}
For example, in the case of relativistic fields, where $\xi^a$
corresponds to a square-integrable solution of the Klein-Gordon
equation defined on a background space-time, this decomposition
corresponds to splitting the fields into positive and negative
frequency parts. Note that $\xi^{a}_{+}$ and $\xi^{a}_{-}$ are
complex eigenstates of the $J^{a}_{\ b}$ operator:
\begin{eqnarray}
J^{a}_{\ b}\xi^{b}_{+} = + \ri  \xi^{a}_{+} \quad {\rm and} \quad
J^{a}_{\ b}\xi^{b}_{-} = - \ri  \xi^{a}_{-}. \label{eq:3.6}
\end{eqnarray}
As a consequence, the Hermitian condition (\ref{eq:3.2}) implies
that two vectors of the same type (for example, a pair of
$J$-positive vectors) are necessarily orthogonal with respect to the
metric $g_{ab}$. Thus, we have $g_{ab}\xi^{a}_{+} \eta^{b}_{+}=0$
for any pair $\xi^a_+,\,\eta^a_+$ of $J$-positive vectors, and
$g_{ab} \xi^{a}_{-}\eta^{b}_{-}=0$ for any pair $\xi^a_-,\,
\eta^a_-$ of $J$-negative vectors.

In the case of a real vector $\xi^a$ it follows from the
decomposition (\ref{eq:x.6}) that $\xi^a_-=\overline{\xi^a_+}$. We
can also split a complex vector into $J$-positive and $J$-negative
parts. However, in the case of the splitting of a complex vector
$\zeta^a= \zeta^{a}_{+} + \zeta^{a}_{-}$ there is no {\it a priori}
relationship between the components $\zeta^a_+$ and $\zeta^a_-$.
That is, if $\zeta^a$ is not real, then
$\zeta^a_-\neq\overline{\zeta^a_+}$. We note that the complex
conjugate of a $J$-positive vector is nevertheless a $J$-negative
vector, and vice versa. More precisely, we have
$\overline{\zeta^a_+}= {\bar\zeta}^a_-$.

In terms of $J$-positive and $J$-negative vectors, the Dirac inner
product (\ref{eq:3.4}) takes a simplified form:
\begin{eqnarray}
\langle\eta|\xi\rangle = \eta^a_- g_{ab}\xi^b_+ . \label{eq:xx.1}
\end{eqnarray}
Equations (\ref{eq:3.4}) and (\ref{eq:xx.1}) are equivalent as we
verify below:
\begin{eqnarray}
\eta^a_- g_{ab}\xi^b_+ &=& \quat (\eta^a+\ri J^a_{\ c}\eta^c) g_{ab}
(\xi^b-\ri J^b_{\ d}\xi^d) \nonumber \\ &=& \quat \left(g_{ab}+
J^c_{\ a}J^d_{\ b}g_{cd}\right)\eta^a\xi^b - \quat \ri
\left(g_{ac}J^c_{\ b}-J^c_{\ a}g_{bc}\right)\eta^a\xi^b \nonumber \\
&=& \half \eta^a (g_{ab}-\ri \Omega_{ab})\xi^b .\label{eq:xx.2}
\end{eqnarray}
Here we have used the relation (\ref{eq:3.2}) and the antisymmetry
of $\Omega_{ab}$.

\section{Quantum-mechanical observables}
\label{s3}

In this section we show how to represent the observables of standard
quantum mechanics in terms of the geometry of the real Hilbert space
${\cal H}$. A quantum-mechanical observable corresponds to a real
symmetric $J$-invariant quadratic form on ${\cal H}$, that is, to a
real tensor $F_{ab}$ satisfying the symmetry condition
\begin{eqnarray}
F_{ab}=F_{ba} \label{eq:z1}
\end{eqnarray}
and the $J$-invariance condition
\begin{eqnarray}
F_{ab}J^a_{\ c}J^b_{\ d} = F_{cd}. \label{eq:z2}
\end{eqnarray}
Note in particular that the quadratic form $g_{ab}$ satisfies
(\ref{eq:z1}) and (\ref{eq:z2}); $g_{ab}$ is the observable
corresponding to the identity. For the expectation value of the
observable $F$ in the state $\xi^a$ we have
\begin{eqnarray}
\frac{\langle\xi|F|\xi\rangle}{\langle\xi|\xi\rangle} =
\frac{F_{ab}\xi^a\xi^b}{g_{ab}\xi^a\xi^b}, \label{eq:z3}
\end{eqnarray}
and more generally given the states $\xi^a$ and $\eta^a$ we have
\begin{eqnarray}
\langle\eta|F|\xi\rangle = \eta_-^a F_{ab}\xi^b_+ . \label{eq:z4}
\end{eqnarray}

The quantum {\sl operator} associated with the observable $F_{ab}$
is obtained by raising one of the indices with the inverse of the
metric:
\begin{eqnarray}
F^a_{\ b} = g^{ac}F_{cb} . \label{eq:z5}
\end{eqnarray}
Then, since $F_{ab}$  is $J$-invariant, it follows that when the
quantum operator $F^a_{\ b}$ acts on a $J$-positive state vector,
the result is another $J$-positive state vector. Alternative ways of
writing (\ref{eq:z4}) are
\begin{eqnarray}
\langle\eta|F|\xi\rangle = \eta_-^a g_{ac}F^c_{\ b} \xi^b_+ =
F^a_{\ c}\eta^c_- g_{ab}\xi^b_+ , \label{eq:z6}
\end{eqnarray}
which express the self-adjointness of $F^a_{\ b}$ with respect to
the Dirac Hermitian inner product.

Let us consider now the symmetries of the Hilbert space ${\cal H}$.
The rotations of ${\cal H}$ around the origin are represented as
orthogonal transformations, which are matrix operations of the form
$\xi^a\to M^a_{\ b}\xi^b$ such that
\begin{eqnarray}
g_{ab}M^a_{\ c}M^b_{\ d} = g_{cd}. \label{eq:z7}
\end{eqnarray}
Such transformations preserve the norm $g_{ab}\xi^a\xi^b$ of the
state $\xi^a$. The unitary group then consists of orthogonal
matrices that also leave the symplectic structure invariant:
\begin{eqnarray}
\Omega_{ab}M^a_{\ c}M^b_{\ d} = \Omega_{cd}. \label{eq:z8}
\end{eqnarray}
In the case of an infinitesimal orthogonal transformation of the
form
\begin{eqnarray}
M^a_{\ b} = \delta^a_{\ b} + \epsilon f^a_{\ b} \label{eq:z9}
\end{eqnarray}
with $\epsilon^2\ll1$, it is straightforward to verify that $f^a_{\
b}$ satisfies
\begin{eqnarray}
g_{ac}f^c_{\ b} + g_{bc}f^c_{\ a} = 0, \label{eq:z10}
\end{eqnarray}
from which we deduce that $f^a_{\ b}$ has the form
\begin{eqnarray}
f^a_{\ b} = g^{ac}f_{cb}, \label{eq:z11}
\end{eqnarray}
where $f_{ab}$ is antisymmetric. Substituting (\ref{eq:z11}) into
(\ref{eq:z9}) and then into (\ref{eq:z8}) shows that for $M^a_{\ b}$
to be a unitary operator it is necessary and sufficient that
$f_{ab}$ be $J$-invariant. This shows that any infinitesimal unitary
transformation can be written in the form
\begin{eqnarray}
M^a_{\ b} = \delta^a_{\ b} + \epsilon J^a_{\ c}F^c_{\ b} ,
\label{eq:z12}
\end{eqnarray}
where $F^a_{\ b}$ is the operator associated with the quantum
observable $F_{ab}$. Conversely, $f_{ab}$ is antisymmetric and
$J$-invariant if and only if it can be expressed in the form
\begin{eqnarray}
f_{ab}=F_{ac}J^c_{\ b}, \label{eq:z125}
\end{eqnarray}
where $F_{ab}$ is symmetric and $J$-invariant. Note that if $F^a_{\
b}$ is proportional to the identity $g^a_{\ b}$, then (\ref{eq:z12})
corresponds to an infinitesimal phase transformation. Also, if
$F^a_{\ b}$ is trace-free, then (\ref{eq:z12}) gives rise to an
infinitesimal special unitary transformation.

Thus, the {\sl operator} $F^a_{\ b}$ is associated with both the
observable $F_{ab}$ as well as the infinitesimal unitary
transformation
\begin{eqnarray}
\xi^a \to \xi^a + \epsilon J^a_{\ b} F^b_{\ c} \xi^c .
\label{eq:z13}
\end{eqnarray}
The complete trajectory of the unitary transformation associated
with the operator $F^a_{\ b}$ can be obtained by exponentiating
(\ref{eq:z13}) and writing
\begin{eqnarray}
\xi^a(t) = \exp\left( t J^b_{\ c} F^c_{\ d}\xi^d \partial_b\right)
\xi^a \Big|_{\xi^a=\xi^a(0)} , \label{eq:z14}
\end{eqnarray}
where $\partial_b=\partial/\partial\xi^b$. The differential operator
in the exponent can be written as
\begin{eqnarray}
J^b_{\ c} F^c_{\ d}\xi^d \partial_b = \half \left( \Omega^{ab}
\partial_b F\right)\partial_a , \label{eq:z15}
\end{eqnarray}
where $F(\xi)=F_{ab}\xi^a\xi^b$. Thus, we see that the quadratic
form $F_{ab}\xi^a\xi^b$ is the generator of a Hamiltonian vector
field $X^a(\xi)=\partial\xi^a/\partial t$ on ${\cal H}$ given by
\begin{eqnarray}
\frac{\partial\xi^a}{\partial t} = \half \Omega^{ab} \partial_b
F(\xi) . \label{eq:z16}
\end{eqnarray}
In other words, the trajectory of the one-parameter family of
unitary transformations associated with the observable $F_{ab}$ is
generated by the Hamiltonian vector field $\frac{1}{2} \Omega^{ab}
\partial_b F(\xi)$. If $H(\xi)=H_{ab}\xi^a\xi^b$ denotes the
quadratic function on ${\cal H}$ associated with the Hamiltonian of
a standard quantum system, then the Schr\"odinger equation can be
written in the form
\begin{eqnarray}
\frac{\partial\xi^a}{\partial t} = \half\Omega^{ab} \partial_b H .
\label{eq:z17}
\end{eqnarray}

We have shown how to describe standard quantum mechanics in terms of
the geometry of a real vector space ${\cal H}$ equipped with a
complex structure $J^a_{\ b}$, a positive-definite quadratic form
$g_{ab}$, and a compatible symplectic structure $\Omega_{ab}$.
Observables are represented by $J$-invariant quadratic forms on
${\cal H}$ and dynamical trajectories are given by the symplectic
vector field on ${\cal H}$ generated by such forms. These structures
are intrinsic to standard quantum theory.

\section{Space-time reflection symmetry}
\label{s4}

In Section \ref{s3} we showed that to describe standard quantum
theory geometrically
it is necessary to introduce a complex structure tensor
$J^a_{\ b}$ on the underlying space ${\cal H}$ of real state
vectors. The remaining structures, namely, the positive definite
quadratic form $g_{ab}$ and the symplectic structure $\Omega_{ab}$,
are then chosen to satisfy the compatibility conditions. In this
section we show how to represent geometrically a
$\mathcal{PT}$-symmetric quantum theory. To do so, we will replace
the metric $g_{ab}$ of standard quantum mechanics by a new quadratic
form $\pi_{ab}$ called {\sl parity}. The novelty of this approach is
that unlike $g_{ab}$, the quadratic form $\pi_{ab}$ is not positive definite.
We will see that the parity operator can only be introduced if the dimension
of the complex vector space of $J$-positive vectors is even.

Recall that in standard quantum mechanics the parity operator
$\pi^a_{\ b}$ represents space reflection and therefore it satisfies
the conditions of an observable, as discussed in the previous
section. This means that $\pi_{ab}=g_{ac}\pi^c_{\ b}$ is required to be
real and symmetric. In addition it must satisfy the $J$-invariance condition
\begin{eqnarray}
\pi_{ab}J^a_{\ c}J^b_{\ d}=\pi_{cd},
\end{eqnarray}
which is equivalent to the commutation relation
\begin{eqnarray}
\pi^a_{\ c}J^c_{\ b} = J^a_{\ c} \pi^c_{\ b}. \label{eq:4.1}
\end{eqnarray}
In addition, the parity operator is required to satisfy the
orthogonality condition
\begin{eqnarray}
g_{ab}\pi^a_{\ c}\pi^c_{\ b}=g_{cd}. \label{eq:4.3}
\end{eqnarray}
As a consequence, the eigenvalues of the parity operator are $\pm1$,
as we now show: Since $\pi_{ab}$ is symmetric, the orthogonality
condition (\ref{eq:4.3}) reads
\begin{eqnarray}
\pi^a_{\ c}\pi^c_{\ b}=\delta^a_{\ b}. \label{eq:4.35}
\end{eqnarray}
Thus, repeated space reflection is equivalent to the identity. If we
diagonalise $\pi^a_{\ b}$, the diagonal entries must be $\pm1$. Once
the number of positive and negative eigenvalues is known, then the
parity operator is unique up to unitary transformations. To see
this, suppose that ${\cal P}$ and ${\cal P}'$ are distinct parity
operators. Because they have the same spectrum, there exists a
unitary transformation that maps one into the other.

In this paper we make the further assumption that the parity
operator is trace-free:
\begin{eqnarray}
\pi^a_{\ a}=0. \label{eq:4.2}
\end{eqnarray}
This condition may not be essential (see Bender \textit{et al}.
2002a), but for simplicity we insist that the condition
(\ref{eq:4.2}) be satisfied so that half of the eigenvalues are $+1$
and the other half of the eigenvalues are $-1$. As a consequence,
$\pi^a_{\ b}$ defines a special unitary operator on the space of
$J$-positive vectors associated with ${\cal H}$. The trace-free
condition also implies that the parity operator can only be defined
if the dimension of the underlying real Hilbert space ${\cal H}$ is
a multiple of four.

To formulate a $\mathcal{PT}$-symmetric quantum theory, we keep the
real Hilbert space ${\cal H}$ with its complex structure $J^a_{\
b}$, and introduce a new inner product on ${\cal H}$ that is defined
in terms of the parity operator. In particular, we introduce a
$\mathcal{PT}$ inner product $\langle\eta\|\xi\rangle$ for the pair
of elements $\xi^a$ and $\eta^a$ in ${\cal H}$ according to
\begin{eqnarray}
\langle\eta\|\xi\rangle = \half \eta^a (\pi_{ab}-\ri \omega_{ab})
\xi^b, \label{eq:4.4}
\end{eqnarray}
where $\omega_{ab}$ is defined by
\begin{eqnarray}
\omega_{ab}=\Omega_{ac}\pi^c_{\ b}. \label{eq:4.44}
\end{eqnarray}
Equivalently, from (\ref{eq:x.3}) we have
\begin{eqnarray}
\omega_{ab}=\pi_{ac}J^c_{\ b}.
\end{eqnarray}
Since $\pi_{ab}$ is an observable in standard quantum mechanics, it
follows that $\omega_{ab}$ is antisymmetric and thus defines a new
symplectic structure on ${\cal H}$ that is compatible with the
complex structure $J^a_{\ b}$. Indeed, one can easily verify the
$J$-invariance condition
\begin{eqnarray}
\omega_{ab}J^a_{\ c}J^b_{\ d}=\omega_{cd}
\end{eqnarray}
associated with the symplectic structure $\omega_{ab}$. We remark
that a Hilbert space endowed with the inner product (\ref{eq:4.4})
is known as the Pontrjagin space (Pontrjagin 1944), the properties
of which have been investigated by Kre{\u\i}n and collaborators
(Kre{\u\i}n 1965, Azizov 1994). For recent work on the relation
between the Kre{\u\i}n space and ${\cal PT}$ symmetry, see Langer \&
Tretter (2004), G\"uenther {\it et al}. (2005), Tanaka (2006), and
Mostafazadeh (2006).

As in standard quantum mechanics, the ${\cal PT}$ inner product
(\ref{eq:4.4}) can be written directly in terms of the $J$-positive
and $J$-negative parts of the vectors $\xi^a$ and $\eta^a$. Recall
in this connection that splitting ${\cal H}$ into $J$-positive and
$J$-negative parts only depends on the complex structure $J^a_{\
b}$, and not on the associated quadratic forms. A short calculation
shows that
\begin{eqnarray}
\langle\eta\|\xi\rangle = \eta^a_- \pi_{ab}\xi^b_+ .
\label{eq:4.45}
\end{eqnarray}
Conversely, from (\ref{eq:4.45}) we get
\begin{eqnarray}
\eta^a_- \pi_{ab}\xi^b_+ = \quat (\eta^a+\ri J^a_{\ c}\eta^c)
\pi_{ab} (\xi^b-\ri J^b_{\ d}\xi^d) \label{eq:4.555}
\end{eqnarray}
by virtue of (\ref{eq:3.5}). Then, using the $J$-invariance of
$\pi_{ab}$ and the antisymmetry of $\omega_{ab}$, as defined by
(\ref{eq:4.44}), we are immediately led back to the inner product
(\ref{eq:4.4}).

We now demonstrate the ${\cal P}$-invariance of $\omega_{ab}$. We
begin by raising the indices of the quadratic form $\pi_{ab}$ using
the metric $g^{ab}$:
\begin{eqnarray}
\pi^{ab}=g^{ac}g^{bd}\pi_{cd}. \label{eq43}
\end{eqnarray}
We then multiply $\pi^{bc}$ by $\pi_{ab}$. Using (\ref{eq:4.35})
we find that
\begin{eqnarray}
\pi_{ab}\pi^{bc} = \delta_a^{\ c}.
\end{eqnarray}
Thus $\pi^{ab}$, as defined in (\ref{eq43}), is the inverse of
$\pi_{ab}$. It is straightforward to verify that the analogously
defined tensor
\begin{eqnarray}
\omega^{ab}=g^{ac}g^{bd}\omega_{cd} \label{eq45}
\end{eqnarray}
satisfies
\begin{eqnarray}
\omega^{ab}=\pi^{ac}\pi^{bd}\omega_{cd} \label{eq46}
\end{eqnarray}
and
\begin{eqnarray}
\omega_{ab}\omega^{bc} = \delta_a^{\ c}. \label{eq47}
\end{eqnarray}
Equation (\ref{eq47}) shows that $\omega^{ab}$ is the inverse of
$\omega_{ab}$. Also, from (\ref{eq45}) and (\ref{eq46}) we deduce
that
\begin{eqnarray}
\pi_a^{\ c}\pi_b^{\ d}\omega_{cd}=\omega_{ab},
\end{eqnarray}
which shows that $\omega_{ab}$ is ${\cal P}$-invariant.

We summarise these results by observing that for the Hermitian
theory we have the compatible system of structures $(J^a_{\
b},g_{ab},\Omega_{ab})$ on ${\cal H}$, whereas the quantum theory
symmetric under space-time reflection comes equipped with the
compatible system of structures $(J^a_{\ b},\pi_{ab},\omega_{ab})$.
The key difference between the two theories is that while $g_{ab}$
is positive definite, $\pi_{ab}$ is indefinite with the split
signature $(+,\cdots,+,-,\cdots,-)$. In particular, given a state
$\xi^a$, its $\mathcal{PT}$ norm, or more precisely its pseudo-norm,
is defined by the expression
\begin{eqnarray}
\langle\xi\|\xi\rangle = \half \pi_{ab} \xi^a \xi^b. \label{eq:4.18}
\end{eqnarray}
This norm can be either positive or negative and in some cases may
even vanish.

To interpret the ${\cal PT}$ norm we establish some identities
concerning the parity splitting of the Hilbert space. Given any real
element $\xi^a$ in ${\cal H}$, we can split it into its positive and
negative parity parts by writing
\begin{eqnarray}
\xi^a = \xi_{\oplus}^a + \xi_{\ominus}^a,
\end{eqnarray}
where
\begin{eqnarray}
\xi_{\oplus}^a = \half (\xi^a+\pi^a_{\ b}\xi^b) \quad {\rm and}
\quad \xi_{\ominus}^a = \half (\xi^a-\pi^a_{\ b}\xi^b).
\end{eqnarray}
These vectors are eigenstates of the parity operator $\pi_{ab}$,
satisfying
\begin{eqnarray}
\pi^a_{\ b}\xi^b_{\oplus} = \xi^a_{\oplus} \quad {\rm and} \quad
\pi^a_{\ b}\xi^b_{\ominus} = -\xi^a_{\ominus}.
\end{eqnarray}
If we write
\begin{eqnarray}
\Pi^a_{\oplus b}=\half(\delta^a_{\ b}+\pi^a_{\ b})\quad{\rm and}
\quad \Pi^a_{\ominus b} = \half (\delta^a_{\ b}-\pi^a_{\ b})
\end{eqnarray}
for the projection operators onto positive and negative parity
eigenstates, then we have
\begin{eqnarray}
\delta^a_{\ b} = \Pi^a_{\oplus b}+ \Pi^a_{\ominus b} \quad {\rm and}
\quad \pi^a_{\ b} = \Pi^a_{\oplus b}- \Pi^a_{\ominus b},
\label{eq:4.24}
\end{eqnarray}
where
\begin{eqnarray}
\Pi^a_{\oplus b}\xi^b=\xi^a_{\oplus} \quad {\rm and} \quad
\Pi^a_{\ominus b}\xi^b=\xi^a_{\ominus}. \label{eq:4.25}
\end{eqnarray}
Because $\pi^a_{\ b}$ and $J^a_{\ b}$ commute, it follows that the
positive parity component of the $J$-positive part of a real vector
$\xi^a$ agrees with the $J$-positive part of the positive parity
part of $\xi^a$, and likewise for other such combinations. This
observation allows us to establish the following result for the
${\cal PT}$ norm:

\begin{guess}
The squared ${\cal PT}$ norm of a state $\xi^a\in{\cal H}$ is
given by the difference between the squared Hermitian norm of the
positive parity part $\xi^a_{\oplus}$ of the state and the squared
Hermitian norm of the negative parity part $\xi^a_{\ominus}$ of
the state:
\begin{eqnarray}
\langle\xi\|\xi\rangle = \langle\xi_{\oplus}|\xi_{\oplus}\rangle
-\langle\xi_{\ominus}|\xi_{\ominus}\rangle. \label{eq:4.26}
\end{eqnarray}
\label{prop1}
\end{guess}

It follows from this proposition that if a measurement of the parity of a state
is more likely to yield a positive result, then its ${\cal PT}$ norm is
positive. Conversely, for a state having more probably negative parity, its
${\cal PT}$ norm is negative. To prove the identity (\ref{eq:4.26}) we
insert (\ref{eq:4.24}) into (\ref{eq:4.18}) and use the relations
(\ref{eq:4.25}) for the parity eigenstates.

We observe finally that if $\xi^a$ and $\eta^a$ are positive and
negative parity states, respectively, then their standard quantum
transition amplitude vanishes:
\begin{eqnarray}
\langle\xi_{\oplus}|\eta_{\ominus}\rangle = 0. \label{eq57}
\end{eqnarray}
We derive (\ref{eq57}) from (\ref{eq:3.4}) by substituting
$\eta^a_{\ominus}$ for $\eta^a$ and $\xi^a_{\oplus}$ for $\xi^a$ and
then using the identities
\begin{eqnarray}
g_{ab}\Pi^a_{\oplus c}\Pi^b_{\ominus d}=0 \quad {\rm and}\quad
\Omega_{ab}\Pi^a_{\oplus c}\Pi^b_{\ominus d}=0.
\end{eqnarray}
The second of these two relations follows from the first because the
$J$-tensor commutes with the parity projection operators.

\section{Observables and symmetries}
\label{s5}

In this section we examine the transformations of ${\cal H}$ that
preserve the ${\cal PT}$ norm $\pi_{ab}\xi^a\xi^b$. Any linear
transformation has the general form $\xi^a\to M^a_{\ b}\xi^b$, and
this transformation preserves the ${\cal PT}$ norm for all
$\xi^a\in{\cal H}$ if and only if
\begin{eqnarray}
\pi_{ab}M^a_{\ c}M^b_{\ d}\xi^c\xi^d = \pi_{ab}\xi^a\xi^b
\label{eq:w1}
\end{eqnarray}
for all $\xi^a$. For an infinitesimal transformation
\begin{eqnarray}
M^a_{\ b} = \delta^a_{\ b} + \epsilon f^a_{\ c}, \label{eq:w2}
\end{eqnarray}
(\ref{eq:w1}) holds to first order in $\epsilon$ if and only if
\begin{eqnarray}
\pi_{ab}f^a_{\ c}\xi^b\xi^c = 0 \label{eq:w3}
\end{eqnarray}
for all $\xi^a$. We deduce that $f^a_{\ b}$ must have the form
\begin{eqnarray}
f^a_{\ b} = \pi^{ac} f_{cb}, \label{eq:w4}
\end{eqnarray}
where $f_{bc}$ is antisymmetric. Here, as in the previous section,
$\pi^{ab}$ denotes the inverse of $\pi_{ab}$ and satisfies
$\pi^{ab}\pi_{bc}=\delta^a_{\ c}$, and we note that $\pi^{ab}$ can
be defined unambiguously in this way without reference to
$g_{ab}$.

To verify (\ref{eq:w4}) we observe that if (\ref{eq:w3}) holds for
all $\xi^a$, then $\pi_{ab}f^b_{\ c}$ must be antisymmetric. Writing
$\pi_{ab}f^b_{\ c}=f_{ac}$, we then obtain (\ref{eq:w4}) by applying
the inverse of $\pi_{ab}$ to each side of the equation. Thus, the
infinitesimal pseudo-orthogonal transformations that preserve the
${\cal PT}$ norm are given by
\begin{eqnarray}
M^a_{\ b} = \delta^a_{\ b} + \epsilon \pi^{ac}f_{cb},
\label{eq:w5}
\end{eqnarray}
where $f_{ab}$ is antisymmetric.

Next we require that the transformations preserve the ${\cal PT}$
symplectic structure $\omega_{ab}$. By virtue of the compatibility
condition, this is equivalent to the condition that the complex
structure is preserved. To first order in $\epsilon$ we have
\begin{eqnarray}
\omega_{ab}M^a_{\ c}M^b_{\ d} = \omega_{cd} + \epsilon \left(
\omega_{ad}\pi^{ae}f_{ec} + \omega_{cb}\pi^{be}f_{ed}\right) .
\label{eq:w6}
\end{eqnarray}
Thus, for $\omega_{ab}$ to be preserved we require that
\begin{eqnarray}
\omega_{ad}\pi^{ae}f_{ec} + \omega_{cb}\pi^{be}f_{ed} =0.
\label{eq:w7}
\end{eqnarray}
However, since $\omega_{ab}=\pi_{ac}J^c_{\ b}$, the condition
(\ref{eq:w7}) implies that $f_{ab}$ is $J$-invariant. Because
$f_{ab}$ is antisymmetric and $J$-invariant, it can be written in
the form
\begin{eqnarray}
f_{ab} = F_{ac}J^c_{\ b} , \label{eq:w8}
\end{eqnarray}
where $F_{ab}$ is a $J$-invariant symmetric quadratic form on
${\cal H}$.

We conclude that the general infinitesimal pseudo-unitary
transformation preserving $\pi_{ab}$ and $\omega_{ab}$ has the form
\begin{eqnarray}
M^a_{\ b} = \delta^a_{\ b} + \epsilon \omega^{ac}F_{cb},
\label{eq:w9}
\end{eqnarray}
where $F_{ab}$ is a standard quantum observable in the sense that it
is symmetric and $J$-invariant. It is interesting to recall equation
(\ref{eq:z125}) and to note that the same $J$-invariant quadratic
forms on ${\cal H}$ appear in standard quantum theory as well as in
${\cal PT}$ symmetric quantum theory.

Following the approach of Section \ref{s3}, we can express the
trajectory of the pseudo-unitary transformation associated with the
operator $F^a_{\ b}=\pi^{ac}F_{cb}$ in the form
\begin{eqnarray}
\xi^a(t) = \exp\left( t \omega^{bc}F_{cd}\xi^d \partial_b\right)
\xi^a \Big|_{\xi^a=\xi^a(0)} ,
\end{eqnarray}
where $\partial_b=\partial/\partial\xi^b$. Therefore, if we write
$F(\xi)=F_{ab}\xi^a\xi^b$ for the quadratic function on ${\cal H}$
associated with a given observable $F_{ab}$, then the dynamical
equation for the corresponding one-parameter family of
pseudo-unitary transformations on ${\cal H}$ preserves the ${\cal
PT}$ inner product, and this equation can be expressed in
Hamiltonian form as
\begin{eqnarray}
\frac{\partial\xi^a}{\partial t} = \half\omega^{ab}\partial_b F .
\end{eqnarray}
This result is analogous to (\ref{eq:z17}) for the case of standard
quantum mechanics.

\section{${\cal PT}$-symmetric Hamiltonian operators}
\label{s6}

In this section we consider observables that are invariant
under space-time reflection symmetry. Specifically, we consider the
properties of $\mathcal{PT}$-symmetric Hamiltonian operators. In
contrast to the Hermiticity condition in conventional quantum
mechanics, here we demand that the Hamiltonian be invariant under
space-time reflection. In ordinary quantum mechanics the Hermiticity
condition on the Hamiltonian operator is that $H^a_{\ b}$ be real,
\begin{eqnarray}
H^a_{\ b} = {\bar H}^a_{\ b} , \label{eq:v1}
\end{eqnarray}
and $J$-invariant,
\begin{eqnarray}
J^a_{\ b}H^b_{\ c}J^c_{\ d} = H^a_{\ d}. \label{eq:v2}
\end{eqnarray}
If a Hamiltonian operator satisfies these conditions, then we say it
is Hermitian. In our discussion of $\mathcal{PT}$-symmetric
Hamiltonian operators, we shall keep the $J$-invariance, but replace
the reality condition by one that has a nice physical
interpretation, namely, invariance under space-time reflection.

In the previous sections we introduced the real vector space ${\cal
H}$ and the complex structure $J^a_{\ b}$ on it. Then we showed that
this structure can be augmented in one of two ways, either by
introducing the positive definite symmetric quadratic form $g_{ab}$
and the associated symplectic structure $\Omega_{ab}$, or by
introducing the split-signature indefinite form $\pi_{ab}$ and the
associated symplectic structure $\omega_{ab}$. In the following, we
will consider either the structure $(J^a_{\ b}, g_{ab},
\Omega_{ab})$ or the structure $(J^a_{\ b},\pi_{ab}, \omega_{ab})$,
or sometimes both. For simplicity of terminology we call the former
the $g$-structure on ${\cal H}$ and the latter the $\pi$-structure
on ${\cal H}$.

We begin by considering those aspects of the ${\cal PT}$-symmetric
theory that arise when we have only the $\pi$-structure on ${\cal
H}$ at our disposal, and we will make no direct use of the parity
operator $\pi^a_{\ b}=g^{ac}\pi_{cb}$ because this involves
$g_{ab}$. We make the following definitions: Suppose that ${\cal H}$
is endowed with a $\pi$-structure and let $H^a_{\ b}$ be a complex
operator on ${\cal H}_{\mathbb C}$ so that $H^a_{\ b}=X^a_{\ b}+\ri
Y^a_{\ b}$, where $X^a_{\ b}$ and $Y^a_{\ b}$ are real. Assume that
$H^a_{\ b}$ is $J$-invariant. Then $H^a_{\ b}$ is said to be
\textit{invariant under space-time reflection}, or ${\cal
PT}$ symmetric, with respect to the given $\pi$-structure if
\begin{eqnarray}
\pi_{bc}{\bar H}^c_{\ d}\pi^{ad} = H^a_{\ b} . \label{eq:v3}
\end{eqnarray}
This relation states that if we take the complex conjugate of the
Hamiltonian followed by a parity transformation, we recover the
original Hamiltonian.

Now we discuss the important notion of a Hermitian form. A tensor
$K_{ab}$ on ${\cal H}_{\mathbb C}$ is said to be a \textit{Hermitian
form} if it is $J$-invariant and satisfies
\begin{eqnarray}
{\bar K}_{ab} = K_{ba} . \label{eq:v4}
\end{eqnarray}
Thus, $K_{ab}$ is a Hermitian form if $K_{ab}=X_{ab}+\ri Y_{ab}$,
where $X_{ab}$ and $Y_{ab}$ are real and $J$-invariant, and $X_{ab}$
is symmetric and $Y_{ab}$ is antisymmetric. In particular,
$g_{ab}-\ri\Omega_{ab}$ and $\pi_{ab}-\ri\omega_{ab}$ are examples
of Hermitian forms. The following proposition is a direct
consequence of these definitions:

\begin{guess}
A Hamiltonian operator $H^a_{\ b}$ is ${\cal PT}$ symmetric with
respect to the $\pi$-structure $(J^a_{\ b},\pi_{ab},\omega_{ab})$
if and only if there exists a Hermitian form $K_{ab}$ such that
\begin{eqnarray}
H^a_{\ b} = \pi^{ac}K_{bc} . \label{eq:v5}
\end{eqnarray}
\label{prop2}
\end{guess}

Proposition~\ref{prop2} demonstrates that the condition of ${\cal PT}$
symmetry on a Hamiltonian is a kind of Hermiticity condition, albeit
not the conventional one. It is possible to characterise ${\cal PT}$
invariance completely without involving any elements of the
$g$-structure on ${\cal H}$. To verify (\ref{eq:v5}) we note that
\begin{eqnarray}
\pi_{bc}{\bar H}^c_{\ d} \pi^{ad} = \pi_{bc}\pi^{ce}{\bar
K}_{de}\pi^{ad} = \delta_b^{\ e}K_{ed}\pi^{ad} = H^a_{\ b}.
\end{eqnarray}

Let us turn now to the analysis of the spectrum of the operator
$H^a_{\ b}$, still keeping within the context of the
$\pi$-structure. Because $H^a_{\ b}$ is complex, we have to admit
the possibility of complex eigenvectors, that is, elements of ${\cal
H}_{\mathbb C}$. The following definition simplifies the exposition:
If $\phi^a$ is an element of ${\cal H}_{\mathbb C}$, then we define
its ${\cal PT}$ norm by the expression $\pi_{ab}\phi^a{\bar\phi}^b$,
which is the sum of the ${\cal PT}$ norms of the real and imaginary
parts of $\phi^a$.

\begin{guess}
If the ${\cal PT}$ norm of an eigenvector of a ${\cal PT}$-symmetric
Hamiltonian is nonvanishing, then the corresponding eigenvalue is
real. \label{prop3}
\end{guess}

{\it Proof}. Suppose that for some possibly complex value of $E$ the
vector $\phi^a$, which may also be complex, satisfies the eigenvalue
equation $H^a_{\ b}\phi^b = E \phi^a$. The complex conjugate of this
equation is ${\bar H}^a_{\ b}{\bar \phi}^b = {\bar E} {\bar
\phi}^a$. Transvecting each side of these equations with $\pi_{ca}$,
we then obtain
\begin{eqnarray}
\pi_{ca}H^a_{\ b}\phi^b = E \pi_{ca}\phi^a \quad {\rm and}\quad
\pi_{ca}{\bar H}^a_{\ b}{\bar \phi}^b = {\bar E} \pi_{ca} {\bar
\phi}^a . \label{eq:v8}
\end{eqnarray}
Therefore, by Proposition~\ref{prop2} we deduce that
\begin{eqnarray}
K_{ab}\phi^b = E \pi_{ab} \phi^b  \label{eq:v9}
\end{eqnarray}
and that
\begin{eqnarray}
{\bar K}_{ab}{\bar \phi}^b = {\bar E} \pi_{ab} {\bar \phi}^b .
\label{eq:v10}
\end{eqnarray}
Because $K_{ab}$ is a Hermitian form, we can replace (\ref{eq:v10})
with the relation
\begin{eqnarray}
K_{ab}{\bar \phi}^b = {\bar E} \pi_{ab} {\bar \phi}^b .
\label{eq:v11}
\end{eqnarray}
If we contract (\ref{eq:v9}) and (\ref{eq:v11}) with ${\bar\phi}^a$
and $\phi^a$, respectively, and subtract, we obtain
\begin{eqnarray}
(E-{\bar E})\pi_{ab}\phi^a{\bar\phi}^b = 0,  \label{eq:v12}
\end{eqnarray}
which establishes Proposition~\ref{prop3}. \hspace*{\fill} $\square$

We conclude that if a ${\cal PT}$-symmetric Hamiltonian has complex
eigenvalues, then the corresponding eigenstates necessarily have a
vanishing ${\cal PT}$ norm. We proceed to augment the vector space
${\cal H}$ with the $g$-structure as well as the $\pi$-structure.
Introducing the $g$-structure allows us to consider the parity
operator $\pi^a_{\ b}$. The condition (\ref{eq:v3}) for the
invariance under space-time reflection can now be written in the
form
\begin{eqnarray}
\pi^a_{\ c}{\bar H}^c_{\ d}\pi^d_{\ b} = H^a_{\ b}. \label{eq:4.8}
\end{eqnarray}
Another way of stating this condition is that the real part of the
Hamiltonian operator has even parity and the imaginary part of the
Hamiltonian has odd parity. Therefore, if we write $H^a_{\ b} =
X^a_{\ b}+\ri Y^a_{\ b}$, where $X^a_{\ b}$ and $Y^a_{\ b}$ are
real, then we have
\begin{eqnarray}
\pi^a_{\ c}X^c_{\ d}\pi^d_{\ b} = X^a_{\ b} \quad {\rm and} \quad
\pi^a_{\ c}Y^c_{\ d}\pi^d_{\ b} = -Y^a_{\ b}. \label{eq:4.88}
\end{eqnarray}
Conversely, any such complex operator is automatically invariant
under space-time reflection.

With the aid of the parity operator $\pi^a_{\ b}$ we are led to the
following observation on the reality of the energy eigenvalues:

\begin{guess}
Let $E$ and $\phi^a$ be an eigenvalue and corresponding eigenstate
of a ${\cal PT}$-symmetric Hamiltonian operator $H^a_{\ b}$. Then,
${\bar E}$ is also an eigenvalue of $H^a_{\ b}$, for which the
associated eigenstate is $\pi^a_{\ b}{\bar\phi}^b$. In particular,
if $\phi^a$ is a simultaneous eigenstate of the ${\cal PT}$
operator, then $E$ is real. \label{prop4}
\end{guess}

{\it Proof}. We start from the eigenvalue equation
\begin{eqnarray}
H^a_{\ b}\phi^b = E \phi^a, \label{eq:4.9}
\end{eqnarray}
where $E$ may or may not be real. Substituting (\ref{eq:4.8})
into
the right side of (\ref{eq:4.9}) gives
\begin{eqnarray}
\pi^a_{\ c}{\bar H}^c_{\ d}\pi^d_{\ b}\phi^b = E \phi^a .
\end{eqnarray}
By taking the complex conjugate, we obtain $\pi^a_{\ c}H^c_{\
d}\pi^d_{\ b}{\bar\phi}^b = {\bar E} {\bar\phi}^a$. We then multiply
on the left by the parity operator and get
\begin{eqnarray}
H^a_{\ b}\pi^b_{\ c}{\bar\phi}^c = {\bar E} \pi^a_{\ b}
{\bar\phi}^b. \label{eq:x5.6}
\end{eqnarray}
Thus, if $\phi^a$ is an energy eigenstate with eigenvalue $E$, then
the state defined by $\pi^a_{\ b} {\bar\phi}^b$ is another energy
eigenstate having eigenvalue ${\bar E}$. If, in addition, the
eigenstate $\phi^a$ is simultaneously an eigenstate of the ${\cal
PT}$ operator, then
\begin{eqnarray}
\pi^a_{\ b}{\bar\phi}^b=\lambda \phi^a , \label{eq:7.180}
\end{eqnarray}
where $\lambda$ is a pure phase. Substituting (\ref{eq:7.180}) into
(\ref{eq:x5.6}) and subtracting the result from (\ref{eq:4.9}), we
get ${\bar E}=E$, which establishes Proposition~\ref{prop4}.
\hspace*{\fill} $\square$

Dorey {\it et al}. (2001a,b) showed that the key condition of
Proposition \ref{prop4}, namely, that $\phi^a$ is a simultaneous
eigenstate of ${\cal PT}$, is in fact valid for the Hamiltonian
$H=p^2+x^2(ix)^\epsilon$ ($\epsilon>0$). When energy eigenstates
$\{\phi^a_n\}$ are not simultaneously eigenstates of the ${\cal PT}$
operator, we say that space-time reflection symmetry is
\textit{broken} (Bender \& Boettcher 1998, Bender \textit{et al}.
1999). In this case, the complex eigenvalues $\{E_n\}$ occur in
complex conjugate pairs. Conversely, if space-time reflection
symmetry is unbroken so that $\{\phi^a_n\}$ are eigenstates of the
${\cal PT}$ operator, then the corresponding energy eigenvalues are
real. In this case, a sufficient (but not necessary) condition for
the orthogonality of the eigenstates can be given:

\begin{guess}
If the eigenstates $\{\phi^a_n\}$ of a ${\cal PT}$-symmetric
Hamiltonian operator $H^a_{\ b}$ are simultaneously eigenstates of
the ${\cal PT}$ operator, then a sufficient condition for the
orthogonality of the eigenstates with respect to the ${\cal PT}$
inner product is that the quadratic form defined by $H_{ab} = g_{ac}
H^c_{\ b}$ is symmetric. \label{prop5}
\end{guess}

{\it Proof}. Consider for $n\neq m$ a pair of eigenvalue equations
$H^a_{\ b}\phi^b_n=E_n\phi^a_n$ and $H^a_{\ b}\phi^b_m =E_m
\phi^a_m$. Transvecting these equations with
$\pi_{ac}{\bar\phi}^c_m$ and $\pi_{ac}{\bar\phi}^c_n$, respectively,
and subtracting the two resulting equations, we obtain
\begin{eqnarray}
{\bar\phi}^c_m\pi_{ca}H^a_{\ b}\phi^b_n -
{\bar\phi}^c_n\pi_{ca}H^a_{\ b}\phi^b_m = \pi_{ab}\left( E_n
\phi^b_n{\bar\phi}^a_m - E_m \phi^b_m{\bar\phi}^a_n\right) .
\label{eq:7.20}
\end{eqnarray}
Now, if the energy eigenstates are simultaneously eigenstates of the
${\cal PT}$ operator so that $\pi^a_{\ b}{\bar\phi}^b_n=\phi^a_n$,
then $\pi_{ab}{\bar\phi}^b_n=g_{ab}\phi^b_n$. Therefore, the left
side of (\ref{eq:7.20}) becomes
\begin{eqnarray}
\phi^c_mg_{ca}H^a_{\ b}\phi^b_n - \phi^c_n g_{ca}H^a_{\ b}\phi^b_m =
H_{cb}\left( \phi^c_m\phi^b_n - \phi^c_n\phi^b_m\right) ,
\end{eqnarray}
where $H_{cb}=g_{ca}H^a_{\ b}$. Therefore, the condition
$H_{cb}=H_{bc}$ is sufficient to ensure that the right side of
(\ref{eq:7.20}) vanishes, which establishes Proposition~\ref{prop5}.
\hspace*{\fill} $\square$

Note that although the symmetric condition on the complex
Hamiltonian $H_{ab}$ is sufficient to ensure the orthogonality of
the eigenstates, it is not a necessary condition.

\section{Construction of a positive inner product}
\label{s7}

In this section we use an additional symmetry operator ${\mathcal
C}$ to construct a positive-definite inner product. It is necessary
to do this because when one formulates quantum mechanics on a
Hilbert space endowed with the structure of space-time reflection
symmetry, one obtains an indefinite metric having a split signature,
where half of the quantum states have positive and the other half
have negative ${\cal PT}$ norm. The split signature arises because
half of the eigenvalues of the parity structure $\pi_{ab}$ are
positive and the other half are negative.

The norm in standard quantum mechanics is closely related to the
probabilistic interpretation of the theory. Therefore, the physical
interpretation of the inner product defined in (\ref{eq:4.4}) is
ambiguous. To remedy this difficulty, Mostafazadeh (2002) and Bender
\textit{et al}. (2002b, 2003) pointed out the existence of a new
symmetry associated with complex non-Hermitian Hamiltonians that are
symmetric under space-time reflection. It was noted that by use of
this symmetry, which carries an interpretation similar to that of
charge conjugation, it is possible to introduce a new inner product
on the vector space ${\cal H}_{\mathbb C}$ spanned by the
eigenstates of ${\cal PT}$-symmetric Hamiltonians in such a way that
all the eigenstates have positive-definite norm. With the aid of
this symmetry the correct probabilistic interpretation of the
quantum theory is restored. Here, we discuss briefly the geometrical
properties of the symmetry associated with the new symmetry operator
$C^a_{\ b}$. We begin by establishing a formula for the ${\cal PT}$
inner product of a pair of energy eigenstates:

\begin{guess}
Suppose that $H^a_{\ b}$ is a ${\cal PT}$-symmetric Hamiltonian
operator whose ${\cal PT}$ symmetry is not broken so that its energy
eigenvalues are real. Let $\{\phi_n^a\}$ denote a set of eigenstates
of $H^a_{\ b}$. Then the ${\cal PT}$ inner product of an arbitrary
pair of energy eigenstates is
\begin{eqnarray}
\langle\phi_m\|\phi_n\rangle = g_{ab}\phi^a_n\phi^b_m .
\label{eq:11.1}
\end{eqnarray}
\label{prop6}
\end{guess}

Recall that the ${\cal PT}$ inner product of a pair of states is
given by $\pi_{ab}\phi^a_n{\bar\phi}^b_m$. Proposition~\ref{prop4}
states that when the ${\cal PT}$ symmetry is unbroken, $\phi^a_n$ is
an eigenstate of the ${\cal PT}$ operator. We thus have
$\pi_{ab}\phi^a_n{\bar\phi}^b_m=g_{ac}\pi^c_{\ b}\phi^a_n
{\bar\phi}^b_m=g_{ac}\phi^a_n\phi^b_m$, which establishes
Proposition~\ref{prop6}. Because the ${\cal PT}$ norms of the energy
eigenstates are real, it follows that the real part of $\phi^a_n$ is
orthogonal to its imaginary part with respect to the quadratic form
$g_{ab}$.

We normalise the energy eigenstates according to the scheme
\begin{eqnarray}
\phi^a_n \to \frac{1}{\sqrt{|g_{ab}\phi^a_n\phi^b_n|}} \phi^a_n ,
\end{eqnarray}
and assume hereafter that $\phi^a_n$ will always be normalised in
this way. It was shown in Section \ref{s4} that half of the
normalised energy eigenstates have positive ${\cal PT}$ norm and
that the remaining half have negative ${\cal PT}$ norm. Without loss
of generality we may order the levels so that
\begin{eqnarray}
g_{ab}\phi^a_m\phi^b_n = (-1)^n \delta_{nm} . \label{eq:11.5}
\end{eqnarray}

With these conventions at hand, we define the new symmetry operator
$C^a_{\ b}$. First, $C^a_{\ b}$ is a ${\cal PT}$-symmetric operator.
This implies that there exists a positive Hermitian form $L_{ab}$
satisfying ${\bar L}_{ab} = L_{ba}$ such that we can write $C^a_{\
b}=\pi^{ac}L_{bc}$. Second, $C^a_{\ b}$ commutes with the
Hamiltonian operator $H^a_{\ b}$. As a consequence, the eigenstates
$\{\phi^a_n\}$ of the Hamiltonian are simultaneous eigenstates of
$C^a_{\ b}$. Third, the eigenvalues of $C^a_{\ b}$ are given by
\begin{eqnarray}
C^a_{\ b}\phi^b_n = (-1)^n \phi^a_n ,
\end{eqnarray}
where $\phi^a_n$ satisfies (\ref{eq:11.5}). In other words, $C^a_{\
b}$ is an operator commuting with the Hamiltonian $H^a_{\ b}$ such
that its eigenvalues are precisely the ${\cal PT}$ norm of the
corresponding eigenstates. Consequently, $C^a_{\ b}$ is involutary,
satisfying $C^a_{\ b}C^b_{\ c}=\delta^a_{\ c}$, and trace-free so
that $C^a_{\ a}=0$.

We remark that in the infinite-dimensional context, it has been
shown that the ${\mathcal C}$ operator admits a position-space
representation of the form (Mostafazadeh 2002, Bender \textit{et
al}. 2002b)
\begin{eqnarray}
{\cal C} = \sum_n \phi_n(x) \phi_n(y) ,
\end{eqnarray}
in contrast with the position-space representation for the parity
operator
\begin{eqnarray}
{\cal P} = \sum_n (-1)^n \phi_n(x) \phi_n(-y) .
\end{eqnarray}
Here $\{\phi_n(x)\}$ denote eigenfunctions of the ${\cal
PT}$-symmetric Hamiltonian.

Having defined the operator $C^a_{\ b}$, we introduce on the vector
space ${\cal H}_{\mathbb C}$ the following inner product: If
$\xi^a,\eta^a\in{\cal H}_{\mathbb C}$, then their quantum-mechanical
inner product $\langle \xi|\eta\rangle$ is defined by
\begin{eqnarray}
\langle\xi|\eta\rangle = g_{ac}C^c_{\ b}\pi^b_{\
d}\eta^a{\bar\xi}^d . \label{eq:9.10}
\end{eqnarray}
With respect to the inner product $\langle\cdot|\cdot\rangle$ we
have
\begin{eqnarray}
\langle\phi_n|\phi_m\rangle = g_{ac}C^c_{\ b}\pi^b_{\ d} \phi^a_m
{\bar\phi}^d_n = g_{ac}C^c_{\ b}\phi^a_m\phi^b_n = (-1)^n
g_{ab}\phi^a_m\phi^b_n = \delta_{nm} .
\end{eqnarray}
Therefore, (\ref{eq:9.10}) defines a positive-definite inner product
between elements of ${\cal H}_{\mathbb C}$. Note that this notation
makes no distinction between the Dirac Hermitian inner product
defined in (\ref{eq:3.4}) and the inner product (\ref{eq:9.10})
defined with respect to ${\cal CPT}$ conjugation. We view
(\ref{eq:9.10}) as a natural extension of (\ref{eq:3.4}) because
when the prescribed Hamiltonian is Hermitian, (\ref{eq:9.10})
reduces to the conventional Dirac inner product (\ref{eq:3.4}).

\section{An explicit two-dimensional construction}
\label{s8}

Consider a quantum-mechanical system of a spin-$\frac{1}{2}$
particle whose Hamiltonian $H$ is a $2\times 2$ complex matrix. We
regard $H$ as an operator that acts on the space of $J$-positive
vectors. The general form of the two-dimensional parity operator
satisfying the properties described above is $\mathcal{P}=
{\boldsymbol\sigma}\cdot{\bf n}$, where ${\bf n}$ is an arbitrary
\textit{real} unit vector and ${\boldsymbol\sigma}$ are the Pauli
matrices. However, because in finite dimensions ${\cal P}$ is
determined uniquely up to unitary transformations, we can set  ${\bf
n}= (1,0,0)$, so that the parity operator is given by
\begin{eqnarray}
\mathcal{P}=\left(\begin{array}{cc} 0 & 1 \cr 1 & 0
\end{array}\right). \label{eq1}
\end{eqnarray}
Based on Wigner's discussion on time reversal in quantum mechanics
(Wigner 1932), we remark that the corresponding operator is
\textsl{antiunitary}. We recall in this connection that a unitary
operator $T$ in conventional quantum mechanics has the
norm-preserving property $\langle\varphi|\psi\rangle=\langle
T\varphi|T\psi\rangle$, whereas if $T$ is antiunitary we have a
`transposed' form of the norm-preserving property
$\langle\varphi|\psi\rangle=\langle T\psi|T\varphi\rangle$ (Wigner
1960a,b). In particular, for a spin system in Hermitian quantum
mechanics the Hamiltonian must be invariant under time
reversal (Morpurgo \& Touschek 1954).

For the present consideration we let time-reversal acting on a
symmetric Hamiltonian be given by complex conjugation. It follows
that a Hamiltonian satisfying the condition ${\cal P}{\bar H}{\cal
P}= H$ can be expressed as
\begin{eqnarray}
H=\left(\begin{array}{cc} r\re^{{\rm i} \theta} & s \cr s &
r\re^{-{\rm i} \theta} \end{array}\right). \label{bbb}
\end{eqnarray}
This is the example considered by Bender \textit{et al}. (2002b,
2003). [A number of other papers have been written on ${\cal
PT}$-symmetric matrix Hamiltonians. See, for example, Znojil (2001),
Mostafazadeh (2002), Weigert (2006), G\"uenther {\it et al}.
(2007).] The Hamiltonian (\ref{bbb}) can alternatively be expressed
in the form
\begin{eqnarray}
H = r\cos\theta {\mathds 1} + \half\omega\,{\boldsymbol\sigma}
\cdot{\bf n}, \label{eq:101}
\end{eqnarray}
where $\omega^2=s^2-r^2\sin^2\theta$ and ${\bf n}=2\omega^{-1}
\left(s,\, 0,\, \ri r \sin\theta\right)$ is a \textit{complex} unit
vector satisfying ${\bf n}\cdot{\bf n}=1$. Therefore, we see that
while a Hermitian Hamiltonian for a spin-$\frac{1}{2}$ particle can
also be written in the form (\ref{eq:101}), the key difference here
is that the unit vector ${\bf n}$ in the case of a ${\cal
PT}$-symmetric system is, in general, complex. This is the sense in
which we are extending quantum mechanics into complex domain.

According to Proposition~\ref{prop2} this operator can
be expressed as a product of the quadratic form representing the
parity operator and a standard Hermitian quadratic form. Thus, we have
\begin{eqnarray}
\left(\begin{array}{cc} r\re^{{\rm i} \theta} & s \cr s &
r\re^{-{\rm i} \theta} \end{array}\right) = \left(\begin{array}{cc}
0 & 1 \cr 1 & 0 \end{array}\right) \left(\begin{array}{cc} s &
r\re^{{\rm i} \theta} \cr r\re^{-{\rm i} \theta} & s
\end{array}\right) . \label{bbbb}
\end{eqnarray}

Although $H$ is a complex matrix, the secular equation for the
eigenvalues of this Hamiltonian is {\sl real} (Bender \textit{et
al}. 2002b). The energy eigenvalues
\begin{eqnarray}
E_\pm = r \cos\theta \pm \sqrt{s^2 -r^2 \sin^2\theta}
\end{eqnarray}
are also real and nondegenerate in the parameter region determined by
\begin{eqnarray}
s^2>r^2 \sin^2\theta. \label{eq:8.5}
\end{eqnarray}
We demand that this inequality be satisfied so that the ${\cal PT}$
symmetry is not broken. If the ${\cal PT}$ symmetry is broken, then
the energy eigenvalues $E_+$ and $E_-$ are complex, and according to
the result of the previous section the ${\cal PT}$ norm of the
corresponding eigenstates must vanish. To verify that the norm
vanishes in this case, we first determine the unnormalised energy
eigenstates and obtain the expression
\begin{eqnarray}
|E_\pm\rangle = \left(\begin{array}{c} 1 \cr -\ri
\frac{r}{s}\sin\theta\pm\sqrt{1-(\frac{r}{s} \sin\theta)^2}
\end{array} \right) .
\end{eqnarray}
Now, if the ${\cal PT}$ symmetry is broken so that $s^2<r^2
\sin^2\theta$, then it follows that the second components of the
vectors $|E_{\pm}\rangle$ are purely imaginary. Recall that if
$|v\rangle=\left({v_1\atop v_2}\right)$ is a two-component vector,
then the application of the $\mathcal{PT}$ operation gives ${\cal
PT} |v\rangle= (\bar{v}_2\ \ \bar{v}_1)$. Therefore, in the broken
symmetry phase, we have
\begin{eqnarray}
\langle E_{\pm}\|E_{\pm}\rangle\Big|_{\rm broken\ {\cal PT}} = 0,
\end{eqnarray}
as claimed.

We now turn to consider the physically interesting situation where
the parameters in the Hamiltonian satisfy (\ref{eq:8.5}) so that the
${\cal PT}$ symmetry is unbroken. In this case we have the
orthogonality condition $\langle E_{\pm}\|E_{\mp}\rangle=0$. The
eigenvectors $|E_{\pm}\rangle$ of the Hamiltonian $H$ are
simultaneously eigenstates of the $\mathcal{PT}$ operator. As
denoted earlier, we can choose the phases of the eigenvectors so
that their eigenvalues under ${\cal PT}$ are all unity. For this
choice of phases these eigenvectors are given by
\begin{eqnarray}
|E_+\rangle = \frac{1}{\sqrt{2\cos\alpha}} \left(
\begin{array}{c} \re^{{\rm i} \alpha/2} \cr
\re^{-{\rm i} \alpha/2} \end{array} \right),\quad |E_-\rangle =
\frac{\ri}{\sqrt{2 \cos\alpha}} \left(\begin{array}{c} \re^{-{\rm i}
\alpha/2} \cr -\re^{{\rm i} \alpha/2}\end{array}\right). \label{ddd}
\end{eqnarray}
Here we have set $\sin\alpha=(r/s)\,\sin\theta$, and the inequality
(\ref{eq:8.5}) for the reality of $E_{\pm}$ ensures that $\alpha$ is
real and that both $st$ and $\cos\alpha$ are positive. It is easy to
verify that these states are also eigenstates of ${\cal PT}$ with
unit eigenvalues.

In conventional Hermitian quantum mechanics the norm is defined in
terms of a Hermitian inner product, which has the form $\langle
u|v\rangle=\bar{u}\cdot v$ and which equals $\bar{u}_1 v_1+
\bar{u}_2 v_2$ in two dimensions. Thus, the norm $\langle v|
v\rangle$ of a vector is positive definite. On the other hand, the
$\mathcal{PT}$ inner product is determined by the $\mathcal{PT}$
conjugation operation $\langle u\|v\rangle=\mathcal{PT} u\cdot v$,
which is $\bar{u}_2 v_1+\bar{u}_1 v_2$ in two dimensions. Note that
$\overline{ \mathcal{PT} u\cdot v} = u\cdot \mathcal{PT} v$. Just as
in the case of the Hermitian norm, the $\mathcal{PT}$ norm $\langle
v\| v\rangle$ is also independent of overall phase. With respect to
the $\mathcal{PT}$ inner product there is an indefinite norm given by
$\langle E_+\|E_+\rangle = +1$ and $\langle E_-\|E_-\rangle = -1$,
as well as the orthogonality conditions $\langle
E_-\|E_+\rangle=\langle E_+\|E_-\rangle=0$. These identities can
easily be verified by use of (\ref{ddd}).

The eigenvectors $|E_\pm\rangle$ are complete in that they span the
two dimensional vector space. The statement of completeness is
embodied in the identity
\begin{eqnarray}
\| E_+\rangle \langle E_+ \| - \| E_-\rangle \langle E_- \| = \left(
\begin{array}{cc} 1 & 0 \cr 0 & 1
\end{array} \right),
\label{eq22}
\end{eqnarray}
where $\| v\rangle \langle v \|$ denotes $|v\rangle\langle{\cal PT}
v|$. Equation (\ref{eq22}) is the $\mathcal{PT}$-symmetric version
of the statement of completeness $|E_+\rangle\langle E_+| +
|E_-\rangle\langle E_-| = {\mathds 1}$ in a Hermitian quantum
theory.

The ${\cal C}$ operator is given by ${\boldsymbol\sigma}\cdot{\bf
n}$, or more specifically by:
\begin{eqnarray}
{\cal C} = \frac{1}{\cos\alpha} \left( \begin{array}{cc} \ri
\sin\alpha & 1 \cr 1 & -\ri\sin\alpha \end{array} \right).
\label{eq22.1}
\end{eqnarray}
Note that $[{\cal C},{\cal PT}]=0$ and $[{\cal P},{\cal T}]=0$
implies ${\cal CPT}={\cal TPC}$. Therefore, if $|v\rangle=
\left({v_1\atop v_2}\right)$ is an arbitrary two-component vector,
we have
\begin{eqnarray}
{\cal CPT}[|v\rangle]&=& {\cal T}\left[ \frac{1}{\cos\alpha} \left(
\begin{array}{cc} 1 & \ri \sin\alpha \cr -\ri\sin\alpha &
1 \end{array} \right) \left({v_1\atop v_2}\right) \right] \nonumber
\\ &=& \frac{1}{\cos\alpha} \Big( \begin{array}{cc} {\bar
v}_1 + \ri {\bar v}_2\sin\alpha \,&\,\,\, {\bar v}_2  - \ri {\bar
v}_1 \sin\alpha \end{array}\Big) = \langle v|. \label{eq22.2}
\end{eqnarray}
It follows that
\begin{eqnarray}
\langle v|u\rangle = \frac{1}{\cos\alpha} \left( {\bar v}_1 u_1 +
{\bar v}_2 u_2 +\ri ( {\bar v}_2 u_1 - {\bar v}_1 u_2)\sin\alpha
\right)
\end{eqnarray}
for the ${\cal CPT}$ inner product of a pair of vectors. In
particular, it is straightforward to verify that $\langle
E_\pm|E_\mp\rangle=0$ and that $\langle E_\pm|E_\pm\rangle=1$. It
also follows that the squared ${\cal CPT}$ norm of an arbitrary
vector $|v\rangle$, given by
\begin{eqnarray}
\langle v|v\rangle = \frac{1}{\cos\alpha} \left( {\bar v}_1 v_1 +
{\bar v}_2 v_2 +\ri ( {\bar v}_2 v_1 - {\bar v}_1 v_2)\sin\alpha
\right),
\end{eqnarray}
is real and positive (because ${\bar v}_2 v_1 - {\bar v}_1 v_2$ is
purely imaginary) and that the constraint (\ref{eq:8.5}) is satisfied.

We observe that by the introduction of the additional structure
${\cal C}$ it is possible to restore a fully consistent quantum
theory of a ${\cal PT}$-symmetric spin-$\frac{1}{2}$ particle
system. It should be noted, however, that the example considered
here is by no means the most general form of a complex extension of
a two-level system in quantum mechanics, as it is evident from the
special form ${\bf n}=2\omega^{-1} \left(s,\, 0,\, \ri r
\sin\theta\right)$ of the unit vector used in (\ref{eq:101}).

\begin{acknowledgments}
We wish to express our gratitude to H.~F.~Jones and R.~F.~Streater
for stimulating discussions. As an Ulam Scholar, CMB receives
financial support from the Center for Nonlinear Studies at the Los
Alamos National Laboratory and he is supported in part by a grant
from the U.S. Department of Energy. DCB gratefully acknowledges
financial support from The Royal Society.
\end{acknowledgments}

\end{document}